\documentclass[preprint,showkeys,amsmath,amssymb,aps]{revtex4}
\usepackage{amsmath}
\usepackage{graphics} 
\usepackage{epsfig}
\usepackage{bm}
\everymath{\displaystyle}
\begin{document}
\title{HAMILTON OPERATOR AND THE SEMICLASSICAL LIMIT FOR SCALAR
PARTICLES IN AN ELECTROMAGNETIC FIELD}
\author{Alexander J. Silenko}
\affiliation{Research Institute for Nuclear Problems of Belarusian
State University, Minsk 220030, Belarus}

\begin {abstract} We successively apply the generalized
Case-Foldy-Feshbach-Villars (CFFV) and the Foldy-Wouthuysen (FW)
transformation to derive the Hamiltonian for relativistic scalar
particles in an electromagnetic field. In contrast to the original
transformation, the generalized CFFV transformation contains an
arbitrary parameter and can be performed for massless particles,
which allows solving the problem of massless particles in an
electromagnetic field. We show that the form of the Hamiltonian in
the FW representation is independent of the arbitrarily chosen
parameter. Compared with the classical Hamiltonian for point
particles, this Hamiltonian contains quantum terms characterizing
the quadrupole coupling of moving particles to the electric field
and the electric and mixed polarizabilities. We obtain the quantum
mechanical and semiclassical equations of motion of massive and
massless particles in an electromagnetic field.
\end{abstract}


\maketitle

\section {Introduction}

The scalar (spinless) particles do not have their proper multipole
moments. Therefore, studying their coupling to an external field
gives an excellent opportunity to compare the conclusions of the
classical and quantum theories. The additional terms in the
quantum mechanical Hamiltonian describe the quantum interaction.

The original equation for spin-0 particles in an external field is
the second-order Klein-Gordon (KG) equation, but if information
about observable quantities is needed, then it is less convenient
than the relativistic wave equation for the Hamiltonian or the
first-order Duffin-Kemmer-Petiau (DKP) equation \cite{D,K,Pe}. The
DKP equation is successfully used in the case of particles with
spins 0 and 1. Here, we transform the Hamiltonian to the diagonal
form characterizing the Foldy-Wouthuysen (FW) representation
\cite{FW}. Using this representation simplifies finding the
eigenvalues and expectation values of operators and deriving the
semiclassical equations of motion (see \cite{JMP}).

Passing from a second-order equation to a first-order equation
must be carefully done. The square root of both sides of an
operator equation is often extracted in this case (see, e.g.,
\cite{AcBt}), but this may lead to inaccuracies \cite{PAN}.
Applying the Case-Foldy-Feshbach-Villars (CFFV) transformation to
the equation for the Hamiltonian \cite{C,Fol,FV} and using the DKP
equation for particles in an external field \cite{I1,I2,I3,I4,I5}
are well-known and well-grounded methods.

The Hamiltonian for relativistic scalar point particles can be
derived by the method proposed in \cite{PAN}. based on
successively applying the CFFV and FW transformations. The FW
transformation for nonrelativistic scalar particles in an
electromagnetic field was introduced in \cite{C,FV,Tanaka}.
We here apply the generalized CFFV transformation, which can also
be used in the massless particle case. We derive an equation for
the Hamiltonian describing the coupling of massive and massless
relativistic scalar particles to an electromagnetic field.

We use the system of units where $\hbar=c=1$В but we explicitly
introduce the constant $\hbar$ in some of the equations in Secs. 5
and 6.

\section {Passing to the first-order equation for scalar particles}

The original KG equation for scalar particles in an
electromagnetic field has the form
\begin{equation} \left[\left(i\frac{\partial}{\partial t}-e\varphi\right)^2-(\bm p-e\bm A)^2-
m^2\right]\psi=0, \label{eq1}\end{equation} where $\varphi$ and
$\bm A$ are the scalar and vector potentials of the
electromagnetic field, $\bm p=-i\nabla,~ e$ and $m$ are the
particle charge and mass, and $\psi$ is a one-component wave
function. This is a second-order equation. One of the main methods
for studying the coupling of scalar particles to an external field
consists in passing to a first-order equation in the time
derivative. This can be done using the CFFV transformation
proposed in \cite{C,Fol,FV}. Its result is a representation of the
wave equation for scalar particles in Hamiltonian form.

We note that a similar transformation to a first-order equation is
also done for spin-1 particles. Although the original relativistic
second-order equations differ significantly for particles with
spins 0 and 1, their transformations have several common features
and can be done using similar methods. For spin-1 particles, such
a transformation was done in \cite{SaTa}. The generalized
Sakata-Taketani transformation for particles with an anomalous
magnetic moment and an electric quadrupole moment was done in
\cite{YB}. In all cases, the result of the transformations is an
equation for an nondiagonal Hamiltonian acting on a bispinor wave
function. The wave functions of the equation for the Hamiltonian
obtained by the CFFV transformation can also be formally regarded
as bispinors. The УspinorsФ for spinless particles are
one-component, and the УbispinorФ wave functions are
two-component.

Although the equations for the nondiagonal Hamiltonians for
particles with integer spin are formally similar to the Dirac
equation, the form of the Hamiltonian and the wave function
normalization differ significantly from those for spin-1/2
particles. At the same time, there is a certain analogy between
the properties of the Hamiltonian and the wave functions for
particles with spins 0 and 1. The wave functions for particles
with spins 0, 1/2, and 1 (the bispinors) can be written in the
general form
\begin{equation} \Psi =\left(\begin{array}{c}
\phi \\ \chi
\end{array}\right),
\label{wavefnc}\end{equation} where $\phi$ and $\chi$ are the
higher- and lower-order spinors. The normalization of wave
functions for particles with spins 0 and 1 is given by
\cite{YB,Davydov}:
$$\int{\Psi^\dagger\rho_3\Psi dV}=1.$$
Here and hereafter, $\rho_i~(i=1,2,3)$ are the Pauli matrices,
whose components act on the spinors $\phi$ and $\chi$:
\begin{equation} \rho_1=\left(\begin{array}{cc}0&1\\1&0\end{array}\right),~~~
\rho_2=\left(\begin{array}{cc}0&-i\\i&0\end{array}\right),~~~
\rho_3\equiv\beta=\left(\begin{array}{cc}1&0\\0&-1\end{array}\right).
\label{eqrho}\end{equation}

The Hamiltonian for particles with spins 0 and 1 is
pseudo-Hermitian or, more precisely, $\beta$-pseudo-Hermitian (see
\cite{Mostafazadeh} and the references therein) and non-Hermitian
in the usual sense. It satisfies the relation \cite{Mostafazadeh}
($\beta^{-1}=\beta$)
$${\cal
H}^\dagger=\beta{\cal H}\beta,$$ which is equivalent to
\cite{YB,Davydov}
\begin{equation} {\cal H}^\ddagger\equiv \rho_3{\cal H}^\dagger\rho_3={\cal H}.
\label{eqghe}\end{equation}

We define [19] the pseudoscalar product of two wave functions
$\Psi_1$ and $\Psi_2$ as the integral \cite{Davydov}
\begin{equation} <\Psi_1|\Psi_2>=\int{\Psi_1^\dagger\rho_3\Psi_2 dV},
\label{eqgin}\end{equation} Then, independently of the original
representation, the pseudoscalar product remains unchanged under a
pseudounitary transformation of the form
$\Psi'_{1,2}=U\Psi_{1,2}$, whose operator has the property
\cite{Davydov}:
\begin{equation} U^\ddagger\equiv \rho_3 U^\dagger\rho_3=U^{-1}.
\label{eqggt}\end{equation}

For scalar particles, an nondiagonal Hamiltonian acting on a
two-component wave function and obtained by transforming the
second-order wave equation (the KG equation) was found by Case
\cite{C}. A method for obtaining the required transformation was
described by Foldy \cite{Fol}. The version of this method proposed
by Feshbach and Villars \cite{FV} is now commonly used. A
representation based on the nondiagonal Hamiltonian given above is
therefore called the CFFV representation.

Using a pseudounitary transformation (such is the FW
transformation), we can transform an nondiagonal Hamiltonian from
the original CFFV representation (for spin-1 particles, the
Sakata-Taketani representation) to the diagonal form (for spin-1
particles, to the block-diagonal form, i.e., to a form diagonal in
two spinors). Such a transformation was done for relativistic free
scalar particles and, in the nonrelativistic limit, for scalar
particles in an electromagnetic field in \cite{C,FV,Tanaka}.


Passing from the second-order KG equation to a first-order
equation using the CFFV transformation is exact. The resulting
equation determines an nondiagonal Hamiltonian acting on the
two-component УbispinorФ wave function (\ref{wavefnc}).

A method for transforming to a first-order equation for free
particles by passing to a two-component wave function $\Psi$ was
given in \cite{Fol}. For particles in an external field, such a
transformation, which is now commonly used, was done by Feshbach
and Villars \cite{FV} (also see \cite{Davydov}). It consists in
introducing wave functions satisfying the conditions \cite{FV,
Davydov}:
\begin{equation} \psi=\phi+\chi, ~~~
i\frac{\partial\psi}{\partial t}-e\varphi\psi= m(\phi-\chi).
\label{eqFVT}\end{equation} In this case, the two-component wave
function has the form
\begin{equation}
\Psi=\frac12\left(\begin{array}{c}
\psi+\frac1m\left[i\frac{\partial\psi}{\partial t}-e\varphi\psi\right] \\
\psi-\frac1m\left[i\frac{\partial\psi}{\partial
t}-e\varphi\psi\right]
\end{array}\right).
\label{eqvf}\end{equation}

It was shown in \cite{Mosta} that the transformation of Eq.
(\ref{eq1}) into a first-order equation in the general case is
done using formulas of form (\ref{eqFVT}),(\ref{eqvf}), where the
particle mass m can be replaced with any nonzero parameter.
Because the CFFV transformation \cite{C,Fol,FV} can be used only
for massive particles, it is interesting to consider the
generalized CFFV transformation \cite{Mosta}, which can also be
used for massless particles. We study the generalized CFFV
transformation in which the mass $m$ can be replaced with an
arbitrary nonzero real constant. The existence of an arbitrary
parameter allows changing the form of the generalized CFFV
transformation in a certain way. As a result, the form of the
intermediate equation with an nondiagonal Hamiltonian (in the
components of the wave function) in the generalized CFFV
representation can also be changed. But, as is shown below, the
final expression for the Hamiltonian in the FW representation is
independent of the form of that intermediate equation.

The generalized transformation that we propose here can also be
used for massless particles. This is very important because
equations for massless particles cannot be obtained from equations
for massive particles by passing to the limit as $m\rightarrow0$
(see \cite{CPPV} and the references therein).

In the approach we use here, the wave functions $\phi$ and $\chi$
are determined by the equations
\begin{equation} \psi=\phi+\chi, ~~~ 
\left(i\frac{\partial}{\partial t}-e\varphi\right)\psi=
N(\phi-\chi), \label{eq3}\end{equation} where $N$ is an arbitrary
nonzero real parameter. If we multiply the last equation by
$i\partial/(\partial t)-e\varphi$ then we can represent Eqs. (9)
in matrix form:
\begin{equation}  i\frac{\partial\Psi}{\partial t}={\cal H}\Psi,
~~~ {\cal
H}=\rho_3\frac{\bm\pi^2+m^2+N^2}{2N}+e\varphi+i\rho_2\frac{\bm\pi^2+m^2-N^2}{2N},
~~~ \label{eq5}\end{equation} where $\bm\pi=\bm p-e\bm
A=-i\nabla-e\bm A$ is the kinetic momentum operator and $\rho_i$
are Pauli matrices (\ref{eqrho}), whose components act on the
corresponding components of the wave function $\Psi$.

An equation equivalent to (\ref{eq5}) was considered in
\cite{Mosta}. Feshbach and Villars [10] considered a special form
of this equation for $N=m$.

\section {The Foldy-Wouthuysen transformation for relativistic particles
in an external field}


The FW representation described in the classic paper \cite{FW}
occupies a special place in relativistic quantum mechanics. In
this representation, the relations between operators are
completely analogous to the relations between the corresponding
classical quantities. The operators in the FW representation have
the same form as in the nonrelativistic quantum mechanics. It is
very important that the coordinate operator $\bm r$ and momentum
operator $\bm p=-i\nabla$ have a very simple form in this
representation. Exactly the FW representation ensures the best
possibility for obtaining the classical limit of relativistic
quantum mechanics \cite{FW,JMP,CMcK}.

The quasidiagonal form (diagonal in two spinors) of the
Hamiltonian is a result of the wave function transformation to a
given representation (the FW transformation). The states with
positive and negative total energy are separated in this case. The
УjitterФ problem (\emph{Zitterbewegung}) does not arise in the FW
representation. The FW transformation radically simplifies the
process of passing to the semiclassical limit, which, in
particular, does not require such a procedure as separating even
parts of the operators.

We use the FW transformation in the one-particle approximation,
where the radiation corrections are not calculated by field theory
methods but are taken into account phenomenologically by including
additional terms in the relativistic wave equations (similarly to
the anomalous magnetic moment \cite{P}). Naturally, the
one-particle approximation can also be used in the relativistic
particle case where the pair creation probability and the losses
due to Bremsstrahlung can be neglected for a given coupling
energy.

The Hamiltonian ${\cal H}$ can be decomposed into operators
commuting and anticommuting with $\rho_3$:
\begin{equation} {\cal H}=\rho_3 {\cal M}+{\cal E}+{\cal
O}, ~~~\rho_3 {\cal M}={\cal M}\rho_3,~~~\rho_3 {\cal E}={\cal
E}\rho_3, ~~~\rho_3 {\cal O}=-{\cal O}\rho_3. \label{eqH}
\end{equation} In the case under study, we have
\begin{equation} {\cal M}=\frac{\bm\pi^2+m^2+N^2}{2N},
~~~ {\cal E}=e\varphi, ~~~ {\cal
O}=i\rho_2\frac{\bm\pi^2+m^2-N^2}{2N}.\label{eq7} \end{equation}

It is very important that the equations determining the
Hamiltonian for particles with spins 0, 1/2, and 1 have the same
form (\ref{eqH}). Therefore, the corresponding expressions for the
operator $U$ transforming the Hamiltonian to the block-diagonal
form formally coincide. This allows using the FW transformation
methods developed for particles with spin 1/2
\cite{FW,JMP,E,B,BD,AB,KT} and spin 1 \cite{JETP2,UNTSMNEpr} for
the scalar particle. But we must take into account that not any
arbitrary diagonalization of the Hamiltonian leads to the FW
representation, as was first shown in \cite{VJ}. As an example, we
mention the Eriksen-Korlsrud transformation \cite{EK}, which
tranforms the Hamiltonian to the block-diagonal form. As proved in
\cite{PRD}, it does not lead to the FW representation even in the
free particle case.

There are several FW transformation methods for nonrelativistic
spin-1/2 particles that allow calculating the relativistic
corrections \cite{FW,E,BD,AB,KT}. The classic method for such a
transformation, proposed in \cite{FW} (also see \cite{BD}) for
${\cal M}=m$ consists in using the operator
\begin{equation}  U=e^{iS}, ~~~S=-\frac{i}{2m}\rho_3{\cal O}. \label{eqq5} \end{equation}

The transformed Hamiltonian can be written as
\begin{equation}  \begin{array}{c}
{\cal H}'={\cal H}+i[S,{\cal H}]+\frac{i^2}{2!}[S,[S,{\cal
H}]]+\frac{i^3}{3!} [S,[S,[S,{\cal H}]]]+\dots\\
-\dot{S}-\frac{i}{2!}[S,\dot{S}]-\frac{i^2}{3!}
[S,[S,\dot{S}]]-\dots, \end{array} \label{eqFW} \end{equation}
where $[\dots,\dots]$ is a commutator. As a result of
transformation (\ref{eqFW}) the Hamiltonian is determined by the
equation \cite{FW,BD}
\begin{equation} {\cal H}'=\rho_3\epsilon+{\cal E}'+{\cal
O}',~~~\rho_3{\cal E}'={\cal E}'\rho_3, ~~~\rho_3{\cal O}'=-{\cal
O}'\rho_3, \label{eqt}\end{equation} where
$\epsilon=\sqrt{m^2+p^2}$ and the odd operator ${\cal O}'$ is now
of the order $O(1/m)$. This transformation can be repeated
multiply to obtain the required accuracy.

But for relativistic particles in an external field, it is
generally rather complicated to pass to the FW representation.
There are serious arguments [32] that the exact solution of this
problem for Dirac particles in an arbitrary external field was
obtained by Eriksen [25]. But the expression found in [25]
contains square roots of matrix operators, which, as a rule, does
not allow obtaining the Hamiltonian explicitly. For relativistic
particles, it is also very difficult to represent it as a power
series in the energy of coupling to the external field. Because
this exact solution is very complicated, it was not used to
perform the FW transformation for relativistic particles in an
external field. Here, such a transformation is done using the
method developed in [5] for spin-1/2 particles. This method allows
finding the relativistic Hamiltonian in the form of a power series
in the potentials of the external field and their derivatives. In
some special cases, this method leads to the exact FW
transformation [5].

The FW transformation was used for nonrelativistic spin-1
particles in an electromagnetic field in [18] and in the
relativistic case in \cite{JETP2,UNTSMNEpr}. It was used also for
nonrelativistic scalar particles in an electromagnetic field in
\cite{C,FV,Tanaka}, and for relativistic particles in [7]. The
nonrelativistic FW transformation was used for a system of two
particles (a spin 0 boson and a spin-1/2 fermion) in [16].

Although the original relativistic wave equations differ
significantly for particles of spins 0 and 1, their
transformations to the FW representation have common features and
are done using similar methods. First, a common feature is that a
preliminary transformation taking the original equations to the
Hamiltonian form is needed. These are the CFFV transformation
\cite{C,Fol,FV} for scalar particles and the Sakata-Taketani
transformation [17] for spin-1 particles. The generalized
Sakata-Taketani transformation for particles with an anomalous
magnetic moment and an electric quadrupole moment was done in
[18]. In all cases, the result is an equation for the nondiagonal
Hamiltonian acting on a bispinor wave function.

The exact FW transformation can be done under the commutation
conditions
\begin{equation}
[{\cal M},{\cal O}]=0,~~~ [{\cal E},{\cal O}]=0, \label{eq8}
\end{equation} and in the case of a stationary external field. In this case, the Hamiltonian
${\cal H}$ is transformed to the block-diagonal form using the
operator \cite{JMP,UNTSMNEpr}
\begin{equation}
U=\frac{\epsilon+{\cal M}+\rho_3{\cal
O}}{\sqrt{2\epsilon(\epsilon+{\cal M})}},~~~
U^{-1}=\frac{\epsilon+{\cal M}-\rho_3{\cal
O}}{\sqrt{2\epsilon(\epsilon+{\cal M})}}, ~~~\epsilon=\sqrt{{\cal
M}^2+{\cal O}^2}. \label{eq9} \end{equation} The transformed
Hamiltonian has the form \cite{JMP,UNTSMNEpr}
\begin{equation}
{\cal H}'=\rho_3\epsilon+ {\cal E}. \label{exct} \end{equation}

In the general case determined by formulas (11) and (12), the
external field is nonstationary, and the operator ${\cal O}$
commutes with ${\cal M}$ but may not commute with ${\cal E}$. We
calculate in the weak-field approximation and assume that the
coupling energy is small compared with the total energy, which
includes the rest energy $mc^2$ and is approximately equal to
$\epsilon$. If the weak-field approximation is used, then the
small dimensionless parameters in which we expand are the ratios
of the terms in the operator of the particle coupling to an
external field (which are proportional to the first powers of the
field potentials and their space and time derivatives) to the
total particle energy. The complete solution of the problem gives
an expression for the Hamiltonian in the FW representation as a
power series in the potentials of the external field and their
derivatives.

The method used here and developed in \cite{JMP,UNTSMNEpr}
consists in the following. First, we perform a pseudounitary
transformation with operator (17) (see \cite{JMP,UNTSMNEpr}).
After this transformation, the Hamiltonian ${\cal H}'$ still
contains odd terms proportional to the potential derivatives and
can be written as
\begin{equation}
{\cal H}'=\rho_3\epsilon+{\cal E}'+{\cal O}',~~~\rho_3{\cal
E}'={\cal E}'\rho_3, ~~~\rho_3{\cal O}'=-{\cal O}'\rho_3,
\label{eq12}\end{equation} where (see \cite{JMP})
\begin{equation}  \begin{array}{c}
{\cal E}'={\cal E}-\frac14\left[\frac{\epsilon+{\cal M}}
{\sqrt{\epsilon(\epsilon+{\cal M})}}\, ,\left[\frac{\epsilon+{\cal
M}}
{\sqrt{\epsilon(\epsilon+{\cal M})}}\, ,{\cal F}\right]\right] \\
+\frac14\left[\frac{\beta{\cal O}} {\sqrt{\epsilon(\epsilon+{\cal
M})}}\, ,\left[\frac{\beta{\cal O}}
{\sqrt{\epsilon(\epsilon+{\cal M})}}\,,{\cal F}\right]\right],\\
{\cal O}'=\frac{\beta{\cal O}}{\sqrt{2\epsilon(\epsilon+{\cal
M})}}\, {\cal F}\frac{\epsilon+{\cal
M}}{\sqrt{2\epsilon(\epsilon+{\cal M})}}- \frac{\epsilon+{\cal
M}}{\sqrt{2\epsilon(\epsilon+{\cal M})}}\,{\cal F}
\frac{\beta{\cal O}}{\sqrt{2\epsilon(\epsilon+{\cal M})}},\\
{\cal F}={\cal E}-i\frac{\partial}{\partial t},
\end{array} \label{eq28} \end{equation}
and $\epsilon$ is determined by formula (17). If the weak-field
approximation is used, then the odd operator ${\cal O}'$ is small
compared with both $\epsilon$ and the original Hamiltonian ${\cal
H}$. The usual scheme of the nonrelativistic FW transformation
\cite{FW,JMP,BD,UNTSMNEpr} can therefore be used at the second
stage. Such a transformation is done using the operator
\begin{equation} U'=\exp{(iS')}, ~~~
S'=-\frac i4\rho_3\left\{{\cal
O}',\frac{1}{\epsilon}\right\}=-\frac
i4\left[\frac{\rho_3}{\epsilon}, {\cal O}'\right], \label{eqtr}
\end{equation} where $\{\dots,\dots\}$ is an anticommutator. The further calculations are the same as in the case of particles with
spins 1/2 and 1 (see \cite{JMP,BD,UNTSMNEpr}). In contrast to
[27], the particle mass in this case must be replaced with the
operator $\epsilon$ noncommuting with the operators ${\cal E}'$
and ${\cal O}'$. If we consider only the leading corrections,
which are proportional to ${{\cal O}'}^2$, i.e., to the second
powers of the field potentials and their space and time
derivatives, then the transformed Hamiltonian becomes \cite{JMP}
\begin{equation}
{\cal H}_{FW}={\cal H}''=\rho_3\epsilon+ {\cal E}'+\frac
{\rho_3}{4}\left\{\frac{1} {\epsilon},{{\cal O}'}^2\right\}.
\label{eqf} \end{equation}

The transformation with operator (21) can be repeated multiply
($S'$ is replaced with $S'',S'''$ and so on) to obtain the
required accuracy.

\section {The Foldy-Wouthuysen transformation for scalar particles}

We perform the FW transformation for relativistic scalar particles
in an electromagnetic field using the method described above. We
assume that the coupling energy is small compared with the total
energy including the rest energy. For the Hamiltonian determined
by formulas (11) and (12), we have
\begin{equation}
\epsilon=\sqrt{m^2+\bm\pi^2}, \label{eq10}
\end{equation}
and the pseudounitary operator of transformation (17) in this case
can be reduced to the form
\begin{equation}
U=\frac{\epsilon+N+\rho_1(\epsilon-N)}{2\sqrt{\epsilon N}}.
\label{eqU}
\end{equation}

As a result of the transformation of the original Hamiltonian
determined by formulas (11) and (12) using operator (24), the
transformed Hamiltonian ${\cal H}'$ also contains odd terms and
has form (19), where
\begin{equation}
{\cal E}'=i\frac{\partial}{\partial
t}+\frac12\left(\sqrt{\epsilon}{\cal
F}\frac{1}{\sqrt{\epsilon}}+\frac{1}{\sqrt{\epsilon}}{\cal
F}\sqrt{\epsilon}\right), ~~~ {\cal
O}'=\frac{\rho_1}{2}\left(\sqrt{\epsilon}{\cal
F}\frac{1}{\sqrt{\epsilon}}-\frac{1}{\sqrt{\epsilon}}{\cal
F}\sqrt{\epsilon}\right),\label{eqHp}\end{equation} and $\epsilon$
is determined by formula (23). The commutation relations can be
used to transform formulas (25) as
\begin{equation}
{\cal E}'={\cal
E}+\frac{1}{2\sqrt{\epsilon}}\left[{\sqrt{\epsilon},
[\sqrt{\epsilon}},{\cal F}]\right]\frac{1}{\sqrt{\epsilon}}, ~~~
{\cal O}'=\rho_1\frac{1}{2\sqrt{\epsilon}}[\epsilon,{\cal
F}]\frac{1}{\sqrt{\epsilon}}. \label{eqHq}\end{equation}

Formulas (23), (25), and (26) are exact for an arbitrary operator
${\cal E}$ and are independent of $N$. This means that the
Hamiltonian in the FW representation, obtained at the next stage
of transformations, is also independent of $N$. The CFFV
representation does not have such a property. Thus, the successive
generalized CFFV and FW transformations again demonstrate the
special role of the FW representation in particle physics, this
time with an example of spin-0 particles.

Approximately (see the general formulas for commutators in
\cite{JMP}),
$$\begin{array}{c}
[\epsilon,{\cal F}]=\frac14\left\{\frac{1}{\epsilon},
[\bm\pi^2,{\cal F}]\right\},~~~ 
\left[{\sqrt{\epsilon}, [\sqrt{\epsilon}},{\cal
F}]\right]=\frac{1}{32}\left\{\frac{1}{\epsilon^3},
\left[\bm\pi^2,[\bm\pi^2,{\cal F}]\right]\right\}.
\end{array} $$

The Hamiltonian in the FW representation, which can be obtained by
approximate formula (22), is equal to
\begin{equation}\begin{array}{c} {\cal H}_{FW}=\rho_3\epsilon+
{\cal E}+\frac{1}{64}\left\{\frac{1}
{\epsilon^4},\left[\bm\pi^2,[\bm\pi^2,{\cal F}]\right]\right\}
+\frac{\rho_3}{64}\left\{\frac{1}
{\epsilon^5},\left([\bm\pi^2,{\cal F}]\right)^2\right\}.
\end{array}\label{eq14}\end{equation}

The above calculations are general and formulas (23)-(27) hold in
the case of the coupling of scalar particles to any external field
whose Hamiltonian is determined by (11) and (12) with arbitrary
${\cal E}$ and $\bm\pi$. In the case of electromagnetic
interaction described by formulas (11) and (12), we have
\begin{equation}\begin{array}{c}
\left[\bm\pi^2,{\cal F}\right]=2ie\bm\pi\cdot\bm E, ~~~
\left[\bm\pi^2,\left[\bm\pi^2,{\cal
F}\right]\right]=4e(\bm\pi\cdot\nabla)(\bm\pi\cdot\bm
E)-4e^2\bm\pi\cdot(\bm E\times\bm H), \\ 
(\bm\pi\cdot\nabla)(\bm\pi\cdot\bm E)\equiv
\pi_i\pi_j\frac{\partial E_j}{\partial x_i}.
\end{array}\label{eq17}\end{equation}

As a result, we obtain the final approximate expression for the
Hamiltonian in the FW representation characterizing the coupling
of scalar point particles to an electromagnetic field:
\begin{equation}\begin{array}{c} {\cal H}_{FW}=\rho_3\epsilon+ e\varphi+
\frac{e}{8\epsilon^4}(\bm\pi\cdot\nabla)(\bm\pi\cdot\bm
E)-\frac{e^2}{8\epsilon^4}\bm\pi\cdot(\bm E\times\bm
H)-\rho_3\frac{e^2}{8\epsilon^5}(\bm\pi\cdot\bm E)^2.
\end{array}\label{eq15}\end{equation}

We note that the successive generalized CFFV and FW
transformations allow deriving a Hamiltonian describing both
massive and massless particles, and this solves the problem of
massless particles in an electromagnetic field. As follows from
(7) and (8), the original transformation \cite{C,Fol,FV} cannot be
done for $m=0$. If the original method \cite{C,Fol,FV} is used,
then the massless particle case cannot be regarded as the limit
case (as $m\rightarrow0$) for massive particles, because the mass
is a multiplier in this approach and it is hence useless to find
the limit as $m\rightarrow0$. Precisely the same situation arises
if we try to use this limit to describe the massless particles by
the DKP equation (see \cite{CPPV}).

In formulas (28) and (29), we do not take the order of operators
into account, bearing in mind the subsequent semiclassical
approximation. In this approach, the condition that the average
commutators of the operators of dynamical variables (coordinates
and momentum) are small compared with the average products of
these operators must be satisfied. The corresponding noncommuting
operators in the quantum mechanical expressions can be rearranged
in this case.

\section {Comparison of the particle properties in classical and quantum
electrodynamics}

The semiclassical transition can be done analogously to [5]. The
above condition that the average commutators of the operators of
coordinates and momentum are small compared with the average
products of these operators is satisfied automatically if the
characteristic length of the domain of the external field
inhomogeneity is significantly greater than the de Broglie
wavelength of the particle: $l\gg\hbar/p$. In this case, the
semiclassical transition is achieved by a trivial replacement of
the operators in the quantum mechanical equations for the
higher-order spinor with the corresponding classical quantities.
The semiclassical Hamiltonian thus obtained is given by
\begin{equation}\begin{array}{c} {\cal H}_s=\epsilon+
e\varphi+\frac{e}{8\epsilon^4}(\bm\pi\cdot\nabla)(\bm\pi\cdot\bm
E)-\frac{e^2}{8\epsilon^4}\bm\pi\cdot(\bm E\times\bm
H)-\frac{e^2}{8\epsilon^5}(\bm\pi\cdot\bm E)^2.
\end{array}\label{eq15n}\end{equation}

The corresponding classical Hamiltonian contains only the first
two terms:
\begin{equation} {\cal H}_c=\epsilon+
e\varphi.\label{eq16}\end{equation}

Although we deal with point particles, the last three terms in
semiclassical Hamiltonian (29) describe properties typical of
classical composite particles. But the Hamiltonian does not
contain any term characterizing the contact (Darwinian)
interaction of rest particles. Moreover, all quantum corrections
become zero for rest particles. This property agrees with the
results obtained in [35]. The derivation of the Hamiltonian
presented there for particles with arbitrary spin and the
subsequent FW transformation for nonrelativistic particles showed
that the appearance of the static contact (Darwinian) interaction
is related to the particle spin. The absence of such an
interaction for spin-0 particles is manifested by the term
characterizing the contact interaction [35] becoming infinite for
$S=0$.

Restoring the constant $\hbar$ in Eq. (29) shows that the last
three terms are proportional to $\hbar^2$:
\begin{equation} {\cal H}_s=\epsilon+
e\varphi+\frac{e\hbar^2}{8\epsilon^4}(\bm\pi\cdot\nabla)(\bm\pi\cdot\bm
E)-\frac{e^2\hbar^2}{8\epsilon^4}\bm\pi\cdot(\bm E\times\bm
H)-\frac{e^2\hbar^2}{8\epsilon^5}(\bm\pi\cdot\bm E)^2.
\label{hbar}\end{equation}

Moving particles have several nonclassical properties. The third
term in (29) describes the quadrupole coupling to the electric
field. A similar term is contained in the Hamiltonian for spin-1/2
particles \cite{JMP,B}. The last term characterizes the electric
polarizability of moving particles, which is also nonzero for
spin-1/2 point particles [26], and the preceding term
characterizes the mixed polarizability. The induced electric
dipole moment is equal to
\begin{equation} \bm d=\frac{\partial{\cal H}_s}{\partial \bm E}=
\frac{e^2\hbar^2}{8\epsilon^4}\bm \pi\times\bm
H-\frac{e^2\hbar^2}{4\epsilon^5}\bm\pi(\bm\pi\cdot\bm
E).\label{eq27}\end{equation}

The induced magnetic dipole moment is determined by the formula
\begin{equation} \bm \mu=\frac{\partial{\cal H}_s}{\partial \bm H}=
-\frac{e^2\hbar^2}{8\epsilon^4}\bm \pi\times\bm
E.\label{eq18}\end{equation}

Hence, the induced electric dipole moment is proportional to the
magnetic field strength, and the induced magnetic dipole moment is
proportional to the electric field strength. These are quantum
properties. The quantum mechanical equation of particle motion has
the form
\begin{equation} \frac{d\bm
\pi}{dt}=\frac{i}{\hbar}[{\cal H}_{FW},\bm \pi]-e\frac{\partial\bm
A}{\partial t}.\label{eqp}\end{equation} In the case under study,
its semiclassical limit is determined by the formula
\begin{equation}\begin{array}{c}  \frac{d\bm\pi}{dt}=e\bm E
+\frac{e}{\epsilon}\left(\bm \pi\times\bm
H\right)+\frac{e^2\hbar^2}{8\epsilon^4}\left[(\bm\pi\cdot\nabla)(\bm
E\times\bm H)-(\bm H\times\nabla)(\bm\pi\cdot\bm
E)\right]\\+\frac{e^3\hbar^2}{8\epsilon^4}\left[H^2\bm E-\bm H(\bm
E\cdot\bm H)\right]-\frac{e^3\hbar^2}{4\epsilon^5}(\bm\pi\cdot\bm
E)(\bm E\times\bm H). \end{array}\label{eql}\end{equation}

The quantum corrections to the classical expression determining
the Lorentz force are very small.

\section {The Duffin-Kemmer-Petiau equation for particles in an
electromagnetic field and the problem of massless particles}

The DKP equation \cite{D,K,Pe} for particles with spins 0 and 1 is
an analogue of the Dirac equation. For free scalar particles, it
has the form
\begin{equation} (\beta^\mu\partial_\mu-m)\Phi=0,
\label{DKe}\end{equation} where
\begin{equation}\begin{array}{c}
\beta^0=\left(\begin{array}{ccccc}0&0&0&0&-1\\0&0&0&0&0\\0&0&0&0&0\\0&0&0&0&0\\1&0&0&0&0
\end{array}\right),~~~
\beta^1=\left(\begin{array}{ccccc}0&0&0&0&0\\0&0&0&0&1\\0&0&0&0&0\\0&0&0&0&0\\0&1&0&0&0
\end{array}\right),\\
\beta^2=\left(\begin{array}{ccccc}0&0&0&0&0\\0&0&0&0&0\\0&0&0&0&1\\0&0&0&0&0\\0&0&1&0&0
\end{array}\right),~~~
\beta^3=\left(\begin{array}{ccccc}0&0&0&0&0\\0&0&0&0&0\\0&0&0&0&0\\0&0&0&0&1\\0&0&0&1&0
\end{array}\right).\end{array}\label{eq21}\end{equation}

The covariant generalization of this equation for spin 0 particles
in an electromagnetic field is \cite{I1,I2,I3}
\begin{equation} (\beta^\mu D_\mu-m)\Phi=0, ~~~ D_\mu=\partial_\mu+ieA_\mu,
\label{DKem}\end{equation} where $A_\mu=(\varphi,-\bm A)$ A) is
the four-dimensional potential of the electromagnetic field. We
write Eq. (39) explicitly as
\begin{equation}\begin{array}{c} -D_0\Phi_5=m\Phi_1, ~~~ D_1\Phi_5=m\Phi_2, ~~~ D_2\Phi_5=m\Phi_3, ~~~
D_3\Phi_5=m\Phi_4,\\
D_0\Phi_1+D_1\Phi_2+D_2\Phi_3+D_3\Phi_4=m\Phi_5.
\end{array}\label{DKeme}\end{equation}

The equivalence of the DKP and KG equations is proved not only for
the electromagnetic interactions \cite{I1,I2,I3} but also for some
others \cite{I2,I3,I5}. Eliminating the components
$\Phi_2,\Phi_3$, and $\Phi_4$ from (\ref{DKeme}), we obtain the
relations
\begin{equation}\begin{array}{c} -D_0\Phi_5=m\Phi_1, ~~~
D_0\Phi_1=\left(m-\frac{D_1^2+D_2^2+D_3^2}{m}\right)\Phi_5.
\end{array}\label{DKemm}\end{equation}

The transformation $\Phi_1=\phi-\chi, ~ \Phi_5=-i(\phi+\chi)$
allows obtaining the equation
\begin{equation}\begin{array}{c}
iD_0\Psi=\left[\rho_3\left(m-\frac{D_1^2+D_2^2+D_3^2}{2m}\right)
-i\rho_2\frac{D_1^2+D_2^2+D_3^2}{2m}\right]\Psi, ~~~
\Psi=\left(\begin{array}{c} \phi \\ \chi \end{array}\right),
\end{array}\label{equiv}\end{equation}
which is equivalent to (10) under the condition that $N=m$.
Eliminating $\Phi_1$ from (\ref{DKemm}), we obtain the original
Eq. (\ref{eq1}).

But DKP equation (39) is inapplicable to massless particles even
in the limit as $m\rightarrow0$ \cite{CPPV}, while the method
proposed here based on the generalized CFFV transformation can
also be used in that case. A modification of the DKP equation that
allows describing free massless particles was found in [22]. The
relativistic wave equation for the Hamiltonian in the FW
representation gives the complete quantum mechanical description
of the coupling of massless particles to an electromagnetic field.
Naturally, this description does not take the quantum field
effects into account. In accordance with formula (29) for the
Hamiltonian, this equation for massless particles has the form
\begin{equation}\begin{array}{c} i\hbar\frac{\partial\Psi_{FW}}{\partial t}=
\left[\rho_3\epsilon+ e\varphi+
\frac{e\hbar^2}{8c^2\epsilon^2}\cdot\frac{\partial E_x}{\partial
x} -\frac{e^2\hbar^2}{8c\epsilon^3}(\bm E\times\bm
H)_x-\rho_3\frac{e^2\hbar^2}{8c^2\epsilon^3}E_x^2\right]\Psi_{FW},
\end{array}\label{eqfin}\end{equation}
where $\epsilon=c\sqrt{\bm\pi^2}$ and the direction of the
particle motion $\bm l=c\bm\pi/\epsilon$ is taken as the $x$ axis.
In the FW representation, we can use only the higher-order spinor
because the lower-order spinor characterizing the states with
negative total energy is equal to zero for real particles. In this
case, the relativistic wave equation becomes
\begin{equation}\begin{array}{c} i\hbar\frac{\partial\phi_{FW}}{\partial t}=
\left[\epsilon+ e\varphi+
\frac{e\hbar^2}{8c^2\epsilon^2}\cdot\frac{\partial E_x}{\partial
x} -\frac{e^2\hbar^2}{8c\epsilon^3}(\bm E\times\bm
H)_x-\frac{e^2\hbar^2}{8c^2\epsilon^3}E_x^2\right]\phi_{FW}.
\end{array}\label{eqfn}\end{equation}

Equations (29), (43), and (44) (just as the original KG equation)
are equivalent to the DKP equation, but their distinguishing
feature is that the quantum corrections to the classical
Hamiltonian are represented explicitly. These corrections are
relativistic and are zero in the static limit. Such corrections
also appear in the quantum mechanical Hamiltonian for spin-1/2
particles \cite{JMP,B}.

\section {Conclusions}

We have transformed the original KG equation for scalar particles
in an electromagnetic field to the Hamiltonian form by
successively using the generalized CFFV and FW transformations. We
derived relativistic formulas for the Hamiltonian and the
semiclassical Hamiltonian. In contrast to the original
transformation, the generalized CFFV transformation contains an
arbitrary parameter and can hence be used for massless particles.
We showed that the form of the Hamiltonian in the FW
representation is independent of the arbitrarily chosen parameter.
We solved the problem of massless particles in an electromagnetic
field. Compared with the classical Hamiltonian for point
particles, the Hamiltonian operator and the semiclassical
Hamiltonian contain quantum terms characterizing the quadrupole
coupling of moving particles to the electric field and the
electric and mixed polarizabilities. These terms are relativistic
and are zero in the static limit. Such quantum corrections to the
classical Hamiltonian also appear in the case of spin-1/2
particles. We obtained quantum mechanical and semiclassical
equations of motion for massive and massless particles in an
electromagnetic field and compared the obtained equations with the
DKP equation.

\textbf{Acknowledgments.} This work is supported by the Belarusian
Foundation for Basic Research.


\begin{thebibliography}{99}
\bibitem{D}
R.J. Duffin, {\em Phys. Rev.}, \textbf{54}:12 (1938), 1114--1114.

\bibitem{K}
N. Kemmer, {\em Proc. Roy. Soc. A},
\textbf{173}:952 (1939), 91--116.
\bibitem{Pe}
G. Petiau, {\em Acad. Roy. de Belg., Classe Sci., Mem. in
8$^\circ$}, \textbf{16}:2 (1936), 1--116.
\bibitem{FW}
L.L. Foldy, S.A. Wouthuysen, {\em Phys. Rev.}, \textbf{78}:1
(1950), 29--36.
\bibitem{JMP}
A.J. Silenko, {\em J. Math. Phys.}, \textbf{44}:7 (2003),
2952--2966.

\bibitem{AcBt} A. Accioly and H. Blas, {\em Phys. Rev. D}, \textbf{66}:6 (2002), 067501;
{\em ibid., Mod. Phys. Lett. A}, \textbf{18}:12 (2003), 867--873.

\bibitem{PAN}
A.J. Silenko, {\em Phys. At. Nucl.}, \textbf{64}:5 (2001),
977--982.

\bibitem{C} K.M. Case, {\em Phys. Rev.}, {\bf 95}:5 (1954), 1323--1328.

\bibitem{Fol}
L.L. Foldy, {\em Phys. Rev.}, {\bf 102}:2 (1956), 568--581.

\bibitem{FV}
H. Feshbach and F. Villars, {\em Rev. Mod. Phys}, \textbf{30}:1
(1958), 24--45.
\bibitem{I1}
M. Nowakowski, {\em Phys. Lett. A}, \textbf{244}:5 (1998),
329--337.
\bibitem{I2}
B.M. Pimentel, V.Ya. Fainberg, {\em Theor. Math. Phys.},
\textbf{124}:3 (2000), 1234--1249.

\bibitem{I3}
V.Ya. Fainberg, B.M. Pimentel, {\em Phys. Lett. A},
\textbf{271}:1--2 (2000), 16--25.
\bibitem{I4}
I.V. Kanatchikov, {\em Rept. Math. Phys.}, \textbf{46}:1--2
(2000), 107--112.
\bibitem{I5}
J.T. Lunardi, B.M. Pimentel, R.G. Teiheira, "Duffin-Kemmer-Petiau
equation in Riemannian space-times", pp. 111--127, in {\em
Geometrical Aspects of Quantum Fields}, ed. by A.A. Bytsenko, A.E.
Golcalves and B.M. Pimentel, World Scientific, 2000.

\bibitem{Tanaka}
T. Tanaka, A. Suzuki, M. Kimura,
{\em Z. Phys. A}, {\bf 353}:1 (1995), 79--85.

\bibitem{SaTa}
M. Taketani and S. Sakata,  {\em Proc. Phys. Math. Soc. Japan}
{\bf 22} (1940), 757--770; {\em ibid., Suppl. Progr. Theor.
Phys.}, 1 (1955), 84--97.

\bibitem{YB}
J.A. Young and S.A. Bludman, {\em Phys. Rev.}, \textbf{131}:5
(1963), 2326--2334.

\bibitem{Davydov}
A.S. Davydov, {\em Quantum Mechanics}, Pergamon Press, New York,
1976.

\bibitem{Mostafazadeh} A. Mostafazadeh,
{\em Czech. J. Phys.}, {\bf 53}:11 (2003), 1079--1084.

\bibitem{Mosta}
A. Mostafazadeh, {\em J. of Phys. A}, {\bf 31}:38 (1998),
7829--7845; {\em ibid., Annals Phys.} {\bf 309}:1 (2004), 1--48.


\bibitem{CPPV}
R. Casana, M. Pazetti, B. M. Pimentel, and J. S. Valverde,
"Pseudoclassical mechanics for the spin 0 and 1 particles",
arXiv:hep-th/0506193 (2005).

\bibitem{CMcK}
J.P. Costella and B.H.J. McKellar, {\em Am. J. Phys.}, {\bf 63}:12
(1995), 1119--1121.

\bibitem{P} W. Pauli, {\em Rev. Mod. Phys.}, {\bf 13}:3 (1941), 203--232.

\bibitem{E} E. Eriksen, {\em Phys. Rev.}, {\bf 111}:3 (1958), 1011--1016.

\bibitem{B} E.I. Blount, {\em Phys. Rev.}, \textbf{128}:5 (1962), 2454--2458.

\bibitem{BD}
J.D. Bjorken and S.D. Drell, {\em Relativistic quantum mechanics},
McGraw-Hill, New York, 1964.

\bibitem{AB}
A.I. Akhiezer and V. B. Berestetskii, {\em Quantum
Electrodynamics}, Interscience, New York, 1965.

\bibitem{KT}
J.G. Korner and G. Thompson, {\em Phys. Lett. B}, {\bf 264}:1--2
(1991), 185--192.

\bibitem{JETP2}
A.J. Silenko, {\em JETP}, \textbf{96}:5 (2003), 775--781.

\bibitem{UNTSMNEpr}
A.J. Silenko, "Analysis of wave equations for spin-1 particles
interacting with an electromagnetic field", arXiv:hep-th/0404074
(2004).

\bibitem{VJ}
E. de Vries, J.E. Jonker, {\em Nucl. Phys. B}, {\bf 6}:3 (1968),
213--225.

\bibitem{EK}
E. Eriksen and M. Korlsrud, {\em Nuovo Cim. Suppl.} {\bf 18}
(1960), 1--39.

\bibitem{PRD} A.J. Silenko and O.V. Teryaev, {\em Phys. Rev. D} {\bf 71}:6 (2005),
064016. 

\bibitem{Nikitin}
A.G. Nikitin, V.I. Fushchich, {\em Theor. Math. Phys.},
\textbf{34}:3 (1978) 203--212.
\end{thebibliography}
\end{document}